\documentclass{iaus}
\usepackage{graphicx}

\title[A search for Type Ia supernova progenitors] %% give here short title %%
{A search for Type Ia supernova progenitors: the central stars of the planetary nebulae NGC
2392 and NGC 6026}

\author[A.~Danehkar, D.\,J.~Frew,  O.~De\,Marco \& Q.\,A.~Parker]   %% give here short author list %%
{A.~Danehkar$^1$, D.\,J.~Frew$^1$, O.~De\,Marco$^1$ \and
Q.\,A.~Parker$^{1,2}$}

\affiliation{$^1$Department of Physics and Astronomy, Macquarie University, Sydney, NSW 2109, Australia \\[\affilskip]
$^2$Australian Astronomical Observatory, PO Box 296, Epping, NSW
1710, Australia\\email: {\tt ashkbiz.danehkar@mq.edu.au;
david.frew@mq.edu.au}}

\pubyear{2011}
\volume{281}  %% insert here IAU Symposium No.
\pagerange{119--126}
 \jname{Binary Paths to the Explosions of type
Ia Supernovae} \editors{Rosanne Di Stefano \& Marina Orio, eds.}
\begin{document}

\maketitle

\begin{abstract}
We use photoionization modeling to assess the binary nature of the central stars of NGC 2392 and NGC 6026.  If they are close binaries, they are potential
Type Ia supernova (SN~Ia) progenitors if the total mass exceeds the Chandrasekhar limit. We show that the nucleus of
NGC 2392 likely has a hot, massive ($\simeq$1\,$M_{\odot}$)
 white dwarf companion, and a total mass of $\sim$1.6\,$M_{\odot}$, making it an especially interesting system.
 The binary mass in NGC 6026 is less, $\sim$1.1\,$M_{\odot}$.  Even though its orbital period is short, it is not considered to be a SN~Ia progenitor.
\end{abstract}

\firstsection % if your document starts with a section,
\section{Introduction}

A proposed pathway for the generation of type Ia supernovae (SN) is
via the merger of two white dwarfs (WDs), the so-called
double-degenerate channel (\cite[Webbink 1984]{Webbink84}). Until
recently, the sample of known double degenerates was limited to
older WDs.  However, recent work has shown that a few planetary
nebula central stars (CSPN) are likely (pre-)double-degenerate
binaries. In this work, we use 1-D (CLOUDY; \cite[Ferland et al.
1998]{Ferland_etal98}) and 3-D (MOCASSIN; \cite[Ercolano et al.
2003]{Ercolano_etal03}) photoionization codes to infer the nature of
the central binary stars in two planetary nebulae (PNe). Constraints
on their component masses will show if the total mass exceeds the
Chandrasekhar limit. If these systems are short-period binaries,
they
 are potential Type Ia SN progenitors.

\section{NGC 2392}

The bright Eskimo nebula (NGC 2392) has a conspicuous
hydrogen-rich CSPN with $T_{\rm eff}$ = 43,000 K (\cite[M\'{e}ndez
et al. 2011]{Mendez_etal11}). However, the surrounding PN has
emission lines of high excitation, such as He \textsc{II}
$\lambda$4686 and [Ne \textsc{v}] $\lambda$3426, which cannot be
produced by the visible star. An additional hot ionizing source must
be present.  To deduce the properties of the companion, we model the
PN using MOCASSIN.  For the model inputs, we used the line
intensities from \cite[Pottasch et al. (2008)]{Pottasch_etal08}, and
adopt a distance of 1.8\,kpc.
%(\cite[Pottasch et al. 2011]{Pottasch_etal11}).
Following \cite[O'Dell et al. (1990)]{ODell_etal90}, the model has a
heterogeneous density distribution, with $n_{\rm e}$ =  3000 and
1300 cm$^{-3}$ for the inner prolate spheroid and outer spherical
zone, respectively (see Fig. 1 in \cite[Danehkar et al.
2011]{Danehkar_etal11}).  Our model outputs agree well with the
observations (see Table~1), and we find that a  hot, massive WD with
$T_{\rm eff} = 250$\,kK is a plausible source of the extra UV
photons. The WD mass from evolutionary tracks is close to
1~$M_{\odot}$, so the total mass may exceed the Chandrasekhar limit.
If the system is a close binary (\cite[M\'{e}ndez et al.
2011]{Mendez_etal11}), the stars may merge within a Hubble time,
making it a potential SN Ia progenitor. It is interesting to note
that \cite[Guerrero et al. (2011)]{Guerrero_etal11} have found the
CSPN to be a hard X-ray source. which may point to mass transfer
between the components. Further observations of this unusual PN and
its central star are urged.

\section{NGC 6026}

Similarly to the Eskimo, spectra show that the excitation class of this elliptical PN is
too high to be the result of photoionization by the
observed CSPN.  This was recently found to be a short-period binary with $P$ = 0.528 days (\cite[Hillwig et
al. 2010]{Hillwig10}).  Based on light-curve modeling, the system consists of a relatively cool O7 star
($T_{\rm eff} = 38$\,kK) and an unseen companion, which is a
hot WD with $T_{\rm eff} = 146$\,kK.   The luminosity and mass of the components, the PN distance, and the nebular abundances were independently determined here by interpolating from a grid of CLOUDY models. The distance is 2.0 $\pm$ 0.5 kpc, and the luminosities and masses %of the cool and hot components are 1500 and 750$L_{\odot}$, and 0.53 and  $0.58~M_{\odot}$ respectively.
of the components are given in Table~1.  Despite the short orbital
period, the total mass is too low for the system to produce a SN Ia.
Lastly, we find that the PN progenitor had sub-solar metallicity,
with [O/H] $\simeq -0.5$ dex.

\begin{table}[tbp]
\caption{Best-fit parameters (left) and observations versus model outputs (right).} \label{table_lines}
\begin{center}
{\small
\begin{tabular}{l|c|c}
\hline \multicolumn{1}{l|} { Parameter } &  {\,NGC 2392} & {\,NGC 6026} \\
\hline
$T_{1}^{\ast}$ $({\rm K})$ & 43\,000 & 37\,500 \\
$L_{1}^{\ast}$ $({\rm L_{\odot}})$& 7\,600 & 1\,500  \\
$M_{1}^{\ast}$ $({\rm M_{\odot}})$& 0.63 & 0.54  \\
$T_{2}^{\ast}$ $({\rm K})$ & 250\,000  &  146\,000 \\
$L_{2}^{\ast}$ $({\rm L_{\odot}})$& 650 & 750  \\
$M_{2}^{\ast}$ $({\rm M_{\odot}})$& 1.0: & 0.60:  \\
${\rm R}_{\rm in} ({\rm pc})$& $0.07$ & ... \\
%${\rm R}_{\rm inner~shell} ({\rm pc})$& $0.07 - 0.09$ & $0.01 - 0.24$ \\
$n_{e\,\rm{(in)}}$ $({\rm cm}^{-3})$& $3000$ & ... \\
${\rm R}_{\rm out}({\rm pc})$& $0.22$ & 0.24\\
%${\rm R}_{\rm outer~shell}({\rm pc})$& $0.09 - 0.22$ & N/A \\
$n_{e\,\rm{(out)}}$ $({\rm cm}^{-3})$& $1300$ & 330\\
$T_{e}$ $({\rm K})$& 14\,500  &  15\,000 \\
$\varepsilon$ & 0.07  &  0.3 \\
${\rm He}/{\rm H}$& $0.08$ & $0.10$ \\
%${\rm C}/{\rm H}$& $3.3(-4)$ & $7.76(-5)$ \\
${\rm N}/{\rm H}$& $1.9(-4)$ & $6.3(-5)$ \\
${\rm O}/{\rm H}$& $2.9(-4)$  & $1.6(-4)$ \\
${\rm Ne}/{\rm H}$& $8.5(-5)$ &  $3.2(-5)$ \\
${\rm S}/{\rm H}$& $7.0(-6)$ &  $5.8(-6)$ \\
${\rm Ar}/{\rm H}$& $2.2(-6)$ &  $7.9(-7)$ \\
\hline
\end{tabular}
}~~~{\small
\begin{tabular}{lc|cc|cc}
\hline \multicolumn{2}{l|} {Object } & \multicolumn{2}{c|} {NGC 2392} & \multicolumn{2}{c} {NGC 6026} \\
\hline Ion & $\lambda$({\AA}) & Obs. & Mod. & Obs. & Mod. \\
\hline $[$Ne \textsc{v}$]$ & 3426
& 4.0 & 2.3 & ... & 19  \\
$[$O \textsc{ii}$]$ & 3727
& 110 & 107 & ...  & 8  \\
$[$Ne \textsc{iii}$]$ & 3869
& 105 & 130 & 34: & 61  \\
H$\gamma$ & 4340
& 47 & 47 & 56 & 48  \\
%$[$O \textsc{iii}$]$ & 4363
%& ... & ... & ... & ...  \\
He \textsc{ii} & 4686
& 37 & 35 & 78 & 89  \\
H$\beta$ & 4861
& 100 & 100 & 100 & 100  \\
$[$O \textsc{iii}$]$  & 5007
& 1150 & 1143 & 709 & 693  \\
$[$N \textsc{ii}$]$  & 5755
& 1.6 & 2.6 & $<$1.0 & 0.4  \\
He \textsc{i}  & 5876
& 7.4 & 7.5 & $<$14 & 2.7  \\
H$\alpha$  & 6563
& 285 & 282 & 286 & 282  \\
$[$N \textsc{ii}$]$  & 6584
& 92 & 129 & 12 & 12  \\
$[$S \textsc{ii}$]$  & 6717
& 6.7 & 3.2 & 2.5: & 6.3  \\
$[$S \textsc{ii}$]$  & 6731
& 8.6 & 4.6 & 2.2:& 5.7  \\
$[$Ar \textsc{iii}$]$  & 7135
& 14 & 12 & 9:: & 6.0  \\
$[$S \textsc{iii}$]$  & 9532 & 91 & 94 & ...  & 60  \\
\hline L(H$\beta$)[erg/s] &$1$E$33$
& 25& 20 & $4.3$  & $5.7$  \\
\hline
\end{tabular}
}
\end{center}
\end{table}

\section*{{\normalsize Acknowledgements}}
AD acknowledges receipt of an MQRES PhD Scholarship and an IAU Travel Grant.

\end{document}